# Election and Subjective Well-Being:
# Evidence from the 2024 U.S. Presidential Election[1]

Dongyoung Kim[2]   Young-Il Albert Kim[3]

Haedong Aiden Rho[4]

[Very Preliminary]

November 2025

Abstract

This paper uses daily Behavioral Risk Factor Surveillance System data to estimate the causal effect of the 2024 U.S. presidential election, a highly competitive race whose outcome resolved lingering uncertainty on election day, on mental-health and life-satisfaction outcomes through a regression discontinuity design. Following the resolution of electoral uncertainty on election day, we find a sharp and persistent post-election decline in subjective well-being, concentrated among female, non-White, urban, and more-educated respondents. These findings reveal an expected-outcome shock, showing that political polarization itself, not electoral surprise, can act as a chronic psychological stressor.

**Codes:** I18, I31

[1] All errors are ours.
[2] Assistant Professor, Department of Economics, Kookmin University, 77 Jeongneung-ro, Seongbuk-gu, Seoul, 02707, South Korea. Email: dongyoung.kim0123@gmail.com.
[3] Corresponding Author, Associate Professor, University of the Fraser Valley. Email: albert.kim@ufv.ca
[4] Assistant Professor, Department of Economics, Universidad del Pacífico. Email: h.rho@up.edu.pe

Recent elections across the world have revealed the deepening intensity of political polarization, fueling ideological, social and psychological divisions. In several cases, political events have even been accompanied by incidents of political violence, reflecting the growing volatility of democratic processes (Das et al., 2023). A growing body of research has shown that unexpected political outcomes—such as the 2016 U.S. presidential election or the Brexit referendum—can trigger short-term declines in subjective well-being. Yet little is known about how the resolution of electoral uncertainty itself—rather than the element of surprise—can affect population wellbeing, even when results fall within expectations. In highly polarized environment, the resolution of electoral uncertainty may impose significant psychological costs on the losing side reflecting the emotional salience of partisan competition rather than pure outcome surprise.

The 2024 U.S. presidential election provides an ideal setting to examine how the resolution of electoral uncertainty affects population well-being. The race remained highly competitive throughout the campaign, with most late-October forecasts showing a statistical tie within the margin of error, and survey data from the Behavioral Risk Factor Surveillance System (BRFSS) show a clear deterioration in mental-health and life-satisfaction indicators immediately after election day. The deterioration followed the resolution of electoral uncertainty and was concentrated among respondents identifying with the losing side, and it persisted for several weeks, contrasting sharply with the pre-election stability and the absence of similar changes in placebo years. These patterns point to a sustained decline in well-being rather than a transient response to news or seasonality.

Taken together, the results may be driven by polarization, rather than the element of surprise, being the primary source of psychological distress. The realization of a political outcome, following the resolution of electoral uncertainty, appears sufficient to produce a measurable welfare loss, particularly among those on the losing side. We interpret this as an outcome-realization shock, in which the resolution of uncertainty and the experience of political loss generate persistent emotional costs. This perspective extends the standard view of election shocks by showing that political polarization can amplify the psychological impact of electoral outcomes, turning the moment of outcome realization into a chronic source of distress for the losing group.

These observations motivate our empirical analysis. To identify the causal impact of the 2024 U.S. presidential election on well-being, we exploit the precise timing of interviews in the BRFSS survey and implement a regression discontinuity design centered on election day, the point at which electoral uncertainty was resolved. This approach isolates short-run discontinuities in reported mental health and life satisfaction while holding constant slow-moving seasonal or demographic factors. By focusing on daily variation in survey timing, we can distinguish the immediate psychological effect of the election from broader macroeconomic or epidemiological trends that might otherwise confound interpretation. The resulting estimates allow us to quantify the magnitude, persistence, and heterogeneity of these effects across key demographic groups (Brookings Institution, 2024; Pew Research Center, 2025).

The main contribution of this paper is to identify and quantify an outcome-realization shock that produces a measurable decline in subjective well-being following the resolution of electoral uncertainty. Prior studies on the 2016 U.S. election provided valuable evidence using BRFSS data (Morey et al., 2021; Yan et al., 2021), relying on multi-month or state-level pre/post comparisons to document broad declines in mental health following unexpected political losses. Our analysis exploits the timing of daily BRFSS interviews using a regression discontinuity design centered on election day, enabling finer temporal resolution and quasi-experimental identification of the immediate causal effect. The pre-election stability and post-election discontinuity reinforce this interpretation. Beyond the primary conceptual contribution, we document substantial heterogeneity across gender, race, education, and urban residence, showing that groups more aligned with the losing side experienced sharper declines in well-being.

We also trace the temporal dynamics of this response. The time-path of subjective well-being around the election, illustrated in Figure 1 and further analyzed in Figure A.2 (impulse-response analysis to be added), suggests that adaptation to the shock was slow and incomplete. The temporal patterns of recovery further connect to the classic hedonic-adaptation framework (Frederick and Loewenstein, 1999), which posits that individuals gradually return toward a baseline level of wellbeing after emotional shocks. Our estimates suggest that this adjustment is considerably slower within a highly polarized political climate: the decline in subjective well-being persists for several weeks. This finding underscores that political polarization can delay emotional adaptation, producing longer-lived welfare effects than typical short-term mood fluctuations. Finally, by linking these findings to broader work on the economic consequences of mental-health shocks, our results suggest that short-term psychological deterioration may translate into longer-term consequences, as observed in changes in birth outcomes following short-term political or social stressors (Dahl et al., 2022; Langer et al., 2024; Oswald et al., 2015).

The paper proceeds as follows. Section 1 explains the context and data. Section 2 presents empirical trends in subjective well-being outcomes before and after the 2024 presidential election. Section 3 examines the impact of the election. Finally, Section 4 concludes the paper.

# 1 Context and Data

The 2024 U.S. presidential election extended and intensified the trajectories seen in the two previous elections. The 2016 election result defied expectations with an outcome that reshaped party coalitions, as Republicans mobilized a new base of working-class, non-college-educated white voters while Democrats became increasingly concentrated among urban, highly educated, female, and minority supporters. This pattern of support rattled the long-standing perception that the Republican supporters as a group primarily composed of high-income, socially conservative elites, signaling a realignment in the social and economic composition of both parties. The 2020 election, held amid the COVID-19 pandemic and widespread social unrest, was marked by a strong partisan alignment. The atmosphere in the 2024 election was held under deeper ideological entrenchment between the two parties and their constituents. The discussions regarding the political choice became increasingly partisan, emphasizing symbolic and identity-based themes rather than concrete policy content.

These developments illustrate a broader and continuing pattern of partisan polarization in the United States, as in many advanced democracies. By 2024, voters had become increasingly anchored to partisan identities, and attitudes toward the opposing side had hardened beyond the realm of ordinary policy disagreement. While Democrats continued to carry women, Black, Hispanic, and urban voters, Republican support within these groups rose substantially in 2024, narrowing the margins that had characterized earlier elections. Social ties based on class, geography, or religion have weakened, leaving political affiliation as a primary marker of identity for many Americans. The result has been a political environment defined less by visions of governance but more by symbolic and identity-related divisions. This evolution, while not unique to the United States, underscores the challenges facing democratic societies as polarization deepens. In such an environment, members of the losing political affiliates and followers may experience heightened stress and a stronger sense of threat, as they may expect the victorious side to enact policies that disadvantage them, and as electoral defeat can be perceived as a symbolic rejection of one's core moral or ideological identity (Abelson et al., 2020).

The 2024 presidential election differed from 2016 in terms of predictability. Whereas the 2016 contest produced a widely unexpected outcome that defied major forecasts, the 2024 race remained highly competitive until the very end, with no clear consensus about the likely winner. By late October, most high-quality national and state polls indicated a close race, with results largely within the margin of error. Although national polls slightly underestimated Republican support, this deviation was well within historical polling variability. In this sense, the 2024 result was not entirely surprising but represented the realization of a closely contested outcome, resolving months of uncertainty.

The primary source of this paper is the 2024 wave of the Behavioral Risk Factor Surveillance System (BRFSS). The BRFSS is a state-based telephone survey that collects health-related information from U.S. adults. A unique advantage of the BRFSS is its temporal coverage, as it surveys a large number of respondents almost every day. Such a large number of respondents enables us to focus on respondents surveyed immediately before and after the 2024 presidential election. In addition, the BRFSS includes subjective well-being information on mental health and life satisfaction, both of which may be influenced by information and expectations about the future, factors potentially affected by the election (Guirola, 2025).

Table 1 presents descriptive statistics for the analysis sample, consisting of respondents surveyed within 50 days of election day. Columns (1) and (2) report the means and standard deviations for the analysis sample, while columns (3) and (4) present the sample means for pre- and post-election periods. Panel A shows the outcome variables: indicators for bad mental health, very bad mental health, satisfied with life, and very satisfied with life.[5] Both mental health and life

---

[5] Indicators for bad mental health and very bad mental health are defined as having 1–13 days and 14–30 days with not good mental health, as pre-classified in the 2024 wave of the BRFSS. Indicators for satisfied with life and very satisfied with life are based on a four-point Likert scale question. The mental health measure is based on responses to the question: "Now thinking about your mental health, which includes stress, depression, and problems

satisfaction deteriorated after the election. In contrast, Panel B shows that individual covariates are similar before and after the election, supporting the validity of comparisons between the two periods. On average, respondents are female, married, white, over age 65, and have a high school diploma. Note that the analysis sample appears similar in covariates when compared with the full 2024 wave of the BRFSS (see Table A.1).

---

with emotions, for how many days during the past 30 days was your mental health not good?" Respondents could report any number of days between 0 and 30. The life satisfaction measure is based on responses to the question: "In general, how satisfied are you with your life?" Respondents chose from a four-point Likert scale with the following options: (1) very satisfied, (2) satisfied, (3) dissatisfied, and (4) very dissatisfied.

## 2 Empirical Trends

We first examine the empirical trends in subjective well-being outcomes before and after the 2024 presidential election. Figure 1 presents the daily averages of the outcome variables. In Panels A and B, we find that both bad mental health and very bad mental health increased on the day of the election. Similarly, Panels C and D show parallel declines in satisfied with life and very satisfied with life on election day. The deterioration in mental health outcomes persisted for at least 50 days following the election. Overall, we observe a consistent pattern of sharp declines in subjective well-being immediately following the election.

As a placebo test, we examine the empirical trends in subjective well-being outcomes one year before, as if the election had occurred on November 6, 2023. If the deterioration in subjective well-being observed on and after election day in 2024 in Figure 1 truly reflects the impact of the election, we should find no discontinuities in the placebo test for 2023. Figure 2 presents the empirical trends for 2023 using the 2023 wave of the BRFSS. It is reassuring that there appears to be no significant jump around the placebo election date in 2023. The null results suggest that the deterioration in subjective well-being in Figure 1 is unlikely to be driven by confounding factors such as seasonality. Note that the distributions of the number of respondents appear similar between the 2024 and 2023 waves of the BRFSS (see Figure A.1).

Table 1: Descriptive Statistics

|  | (1) | (2) | (3) | (4) |
|---|---|---|---|---|
|  | Analysis Sample | | Pre | Post |
|  | Mean | Std Dev |  |  |
| *Panel A. Outcome Variable* | | | | |
| Bad Mental Health | 0.402 | 0.490 | 0.399 | 0.405 |
| Very Bad Mental Health | 0.134 | 0.341 | 0.133 | 0.136 |
| Satisfied with Life | 0.944 | 0.230 | 0.945 | 0.942 |
| Very Satisfied with Life | 0.450 | 0.497 | 0.451 | 0.449 |
| *Panel B. Covariates* | | | | |
| Male | 0.476 | 0.499 | 0.476 | 0.476 |
| Age 18 to 24 | 0.056 | 0.230 | 0.056 | 0.056 |
| Age 25 to 34 | 0.099 | 0.298 | 0.101 | 0.096 |
| Age 35 to 44 | 0.125 | 0.331 | 0.125 | 0.126 |
| Age 45 to 54 | 0.137 | 0.344 | 0.137 | 0.136 |
| Age 55 to 64 | 0.173 | 0.379 | 0.172 | 0.175 |
| Age 65 and over | 0.410 | 0.492 | 0.408 | 0.411 |
| Married | 0.524 | 0.499 | 0.520 | 0.527 |
| High School Diploma | 0.949 | 0.219 | 0.947 | 0.951 |
| College Diploma | 0.434 | 0.496 | 0.435 | 0.434 |
| Employed | 0.498 | 0.500 | 0.502 | 0.494 |
| White | 0.765 | 0.424 | 0.761 | 0.769 |
| Urban | 0.861 | 0.346 | 0.865 | 0.858 |
| Sample Size | 51,740 | | 24,954 | 26,786 |

Note: This table presents the descriptive statistics of the analysis sample. Columns (1) and (2) show the means and standard deviations for the entire analysis sample, while columns (3) and (4) report the means for the pre- and postelection periods under consideration.

Figure 1: Empirical Trends of Subjective Well-Being Outcomes

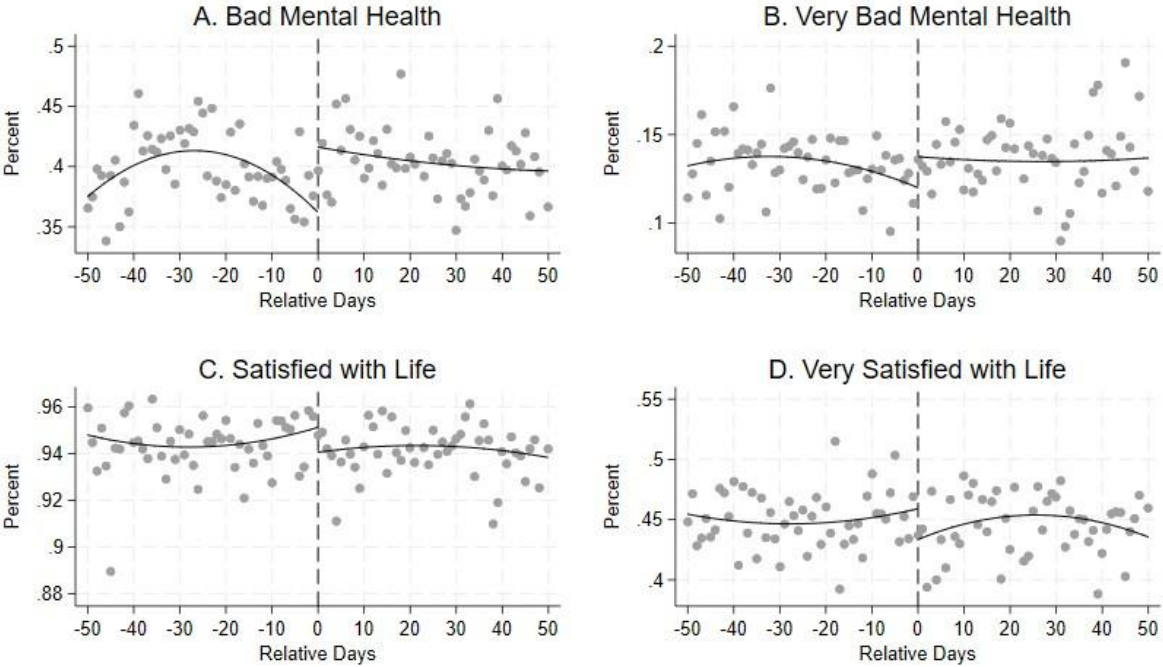

Note: This figure displays the empirical regression discontinuity plot on the relative days from election day. The bandwidth is set to 50 days, and the fitted values are estimated using a local quadratic polynomial. The 2024 wave of the Behavioral Risk Factor Surveillance System is used for the figure.

Figure 2: Placebo Tests on Subjective Well-Being Outcomes

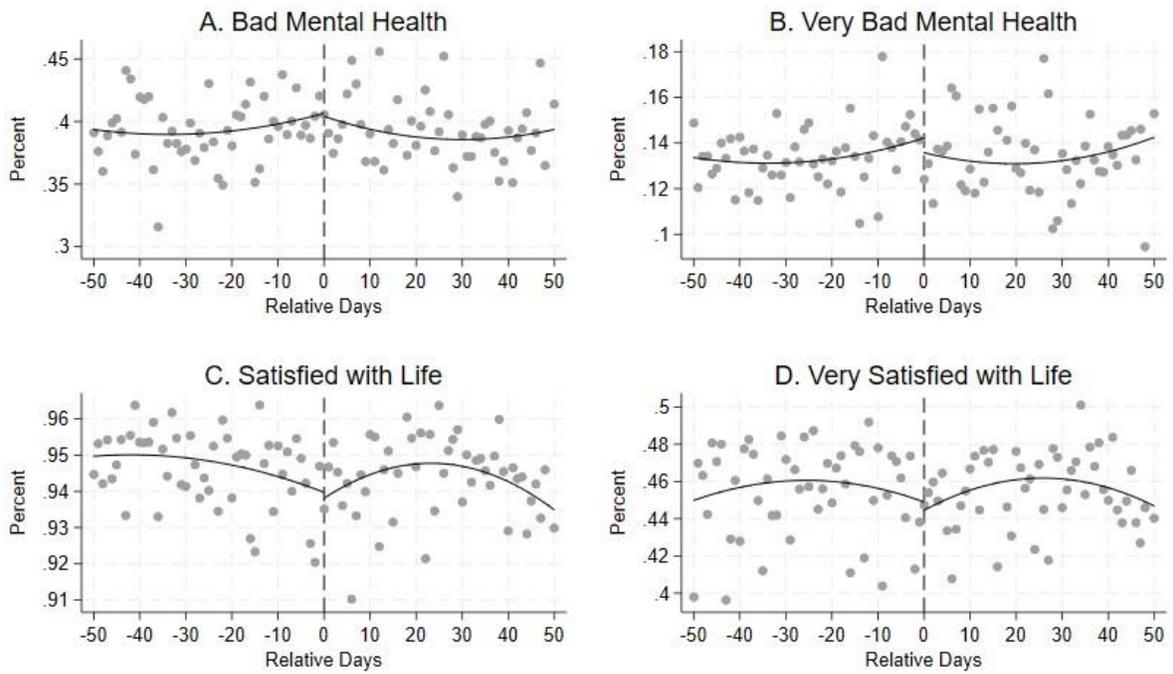

Note: This figure displays the empirical regression discontinuity plot on the relative days from November 6, 2023, as if the election occurred a year before. The bandwidth is set to 50 days, and the fitted values are estimated using a local quadratic polynomial. The 2023 wave of the Behavioral Risk Factor Surveillance System is used for the figure.

# 3 Impact of Election

## 3.1 Baseline Impact

To estimate the subjective well-being effects of the 2024 presidential election, we employ a regression discontinuity in time model:

$$Y_{it} = \beta Post_t + f(Score_t) + X_i\Gamma + \delta_s + \varepsilon_{it} \quad [1]$$

where $Y$ represents subjective well-being outcomes. $Post$ is equal to one if the respondent was surveyed after election day, $Score$ denotes the daily running variable for treatment group, and $\delta_s$ denotes the state fixed effects. The function $f(Score_t)$ is approximated using local quadratic polynomials on either side of the cutoff, weighted by a triangular kernel. $X$ is a vector of socioeconomic covariates.[6] The main coefficient of interest, $\beta$, summarizes the average impact of the election. We use a fixed symmetric bandwidth of 50 days on each side of the cutoff, and the standard errors are clustered at the state level.

Table 2 presents the estimates from equation [1]. Consistent with Figure 1, subjective wellbeing outcomes deteriorated significantly. Panel A reports the results for the baseline model with no controls. In column (1), the proportion of individuals with bad mental health, reporting at least one day of bad mental health in the past 30 days, increased significantly. Similarly, columns (3) and (4) indicate that life satisfaction declined significantly. Although very bad mental health, defined as experiencing at least 14 days of poor mental health in the past 30 days, did not increase significantly, this measure may respond more slowly to the shock. The estimated impact of the election remains robust to the inclusion of additional controls in Panels B and C.

---

[6] This paper controls for the covariates listed in Table 1: male, age group indicators (18–24, 35–44, 45–54, 55–64, and 65+, with 25–34 as the reference group), and education indicators (high school diploma and college diploma).

The magnitude of the estimated impact of the election is sizable. Based on the estimates in Panel C, the incidence of bad mental health increased by 4.2 percentage points, corresponding to 10.5% of the pre-period average. A 1.4 percentage point increase in very bad mental health also represents 11.3% of the pre-period average, although it is statistically insignificant. The 1.0 and 2.8 percentage point declines in being satisfied and very satisfied with life correspond to 1.1% and 6.2% of their respective pre-period averages. Overall, the consistent pattern suggests a meaningful deterioration in subjective well-being outcomes.

Table 2: Baseline Results

|  | (1) | (2) | (3) | (4) |
|---|---|---|---|---|
|  | Bad Mental Health | Very Bad Mental Health | Satisfied with Life | Very Satisfied with Life |
| *Panel A. No Control* | | | | |
| Election | 0.045** | 0.015 | -0.012* | -0.032* |
|  | (0.020) | (0.011) | (0.007) | (0.017) |
| Sample Size | 51,740 | 51,740 | 51,740 | 51,740 |
| *Panel B. Age, Gender, and Race Controlled* | | | | |
| Election | 0.040*** | 0.014 | -0.012* | -0.033** |
|  | (0.015) | (0.009) | (0.007) | (0.016) |
| Sample Size | 51,740 | 51,740 | 51,740 | 51,740 |
| *Panel C. Fully Controlled* | | | | |
| Election | 0.042*** | 0.014* | -0.010* | -0.028** |
|  | (0.012) | (0.007) | (0.006) | (0.012) |
| Sample Size | 51,740 | 51,740 | 51,740 | 51,740 |
| Sample Mean | 0.399 | 0.133 | 0.945 | 0.451 |

Note: Panel A displays the regression discontinuity (RD) estimates from equation [1] without any covariates. Panel B displays the RD estimates with age, gender, and race controlled. Panel C displays the RD estimates with the full set of covariates which include age, gender, and race, marital status, education, employment status, and the state fixed effects. Standard errors are clustered at the state level. The 2024 wave of the Behavioral Risk Factor Surveillance System is used for the table. *, **, and *** denote statistical significance at the 90 percent, 95 percent, and 99 percent levels, respectively.

## 3.2 Heterogeneity

We further explore the heterogeneous impact of the election across individuals' demographic and socioeconomic characteristics. Specifically, we examine heterogeneity by gender, race, urban status, and educational attainment by stratifying the sample along these dimensions. Male, White, rural residents, and those without a high school diploma are generally more likely to support the Republican Party (Brookings Institution, 2024; Pew Research Center, 2025). Therefore, if the patterns observed in Figure 1 and Table 2 reflect effects of the election, the psychological toll would likely have been greater among female, non-White, urban residents, and those with a high school diploma.

Table 3 presents the estimated effects of the election from equation [1] on subsamples across gender, race, urban status, and high school education. We observe that the effects are generally stronger among female, non-White, urban residents, and those with a high school diploma. The results are consistent with known patterns of political preferences by demographic group (Brookings Institution, 2024; Pew Research Center, 2025). This evidence supports the interpretation that the observed decline in subjective well-being reflects the impact of the election. Table A.2 also considers other demographic characteristics such as age, marital status, college education, and employment status, and finds relatively smaller heterogeneity.

Table 3: Heterogeneity by Gender, Race, Urban Status, and High School Education

|  | (1) | (2) | (3) | (4) | (5) | (6) | (7) | (8) |
|---|---|---|---|---|---|---|---|---|
|  | Male | Female | White | Non-White | Urban | Rural | No High School | High School |
| *Panel A. Bad Mental Health* | | | | | | | | |
| Election | 0.032* | 0.050** | 0.032*** | 0.061** | 0.050*** | -0.019 | -0.037 | 0.045*** |
|  | (0.017) | (0.020) | (0.012) | (0.027) | (0.011) | (0.031) | (0.045) | (0.012) |
| *Panel B. Very Bad Mental Health* | | | | | | | | |
| Election | -0.005 | 0.030*** | 0.000 | 0.056*** | 0.018** | -0.013 | 0.019 | 0.014* |
|  | (0.008) | (0.011) | (0.009) | (0.016) | (0.008) | (0.016) | (0.042) | (0.008) |
| *Panel C. Satisfied with Life* | | | | | | | | |
| Election | 0.000 | -0.020** | -0.013** | -0.001 | -0.010* | -0.011 | -0.019 | -0.010* |
|  | (0.009) | (0.008) | (0.006) | (0.012) | (0.006) | (0.014) | (0.035) | (0.006) |
| *Panel D. Very Satisfied with Life* | | | | | | | | |
| Election | -0.021 | -0.033** | -0.023* | -0.036 | -0.025* | -0.048 | -0.022 | -0.028** |
|  | (0.014) | (0.017) | (0.013) | (0.023) | (0.013) | (0.031) | (0.059) | (0.012) |
| Sample Size | 24,622 | 27,118 | 39,574 | 12,166 | 44,560 | 7,180 | 2,624 | 49,116 |

Note: This table displays subgroup analyses by each group with the full set of covariates which include age, gender, and race, marital status, education, employment status, and the state fixed effects. Standard errors in parentheses are clustered at the state level. The 2024 wave of the Behavioral Risk Factor Surveillance System is used for the table. *, **, and *** denote statistical significance at the 90 percent, 95 percent, and 99 percent levels, respectively.

# 4 Conclusion

This paper provides causal evidence that the 2024 U.S. presidential election, one of the most polarized and contentious electoral events in recent history, generated a significant negative impact on population subjective well-being, despite that the election result was not against the pre-election prediction. Using a regression discontinuity design, we find a precipitous deterioration in mental health and life satisfaction immediately following the election. The effects are concentrated among female, non-White, urban respondents, and those with higher education levels. Given the absence of similar changes in the placebo test from the prior year, the evidence may support the interpretation that the election was the primary driver of the observed decline. This paper contributes to the literature by documenting the significant impact of the election even though the election outcome was largely expected. The analysis further reveals a sharp and persistent post-election decline in subjective well-being, suggesting that electoral stress can have lasting psychological consequences and potentially spill over into broader health and social outcomes.

The significant and sizable effects of the election highlight how political polarization and identity-based contestation can generate meaningful welfare losses even in the absence of material shocks. These findings suggest that democratic processes, when conducted under heightened social tension, may impose subjective well-being costs on the electorate. From a policy perspective, the results underscore the importance of fostering electoral environments that minimize hostility and misinformation. Strengthening social capital, promoting balanced media exposure, and facilitating cross-partisan dialogue may help mitigate the well-being costs associated with electoral cycles. Exploring whether short-term declines in subjective well-being have cumulative or spillover consequences for health, productivity, and social capital would be a fruitful direction for future research. In addition, understanding how institutions mediate these effects could help clarify the broader societal implications of political polarization.

# A     Appendix – For Online Publication Only

## A.1     Appendix Figures and Tables

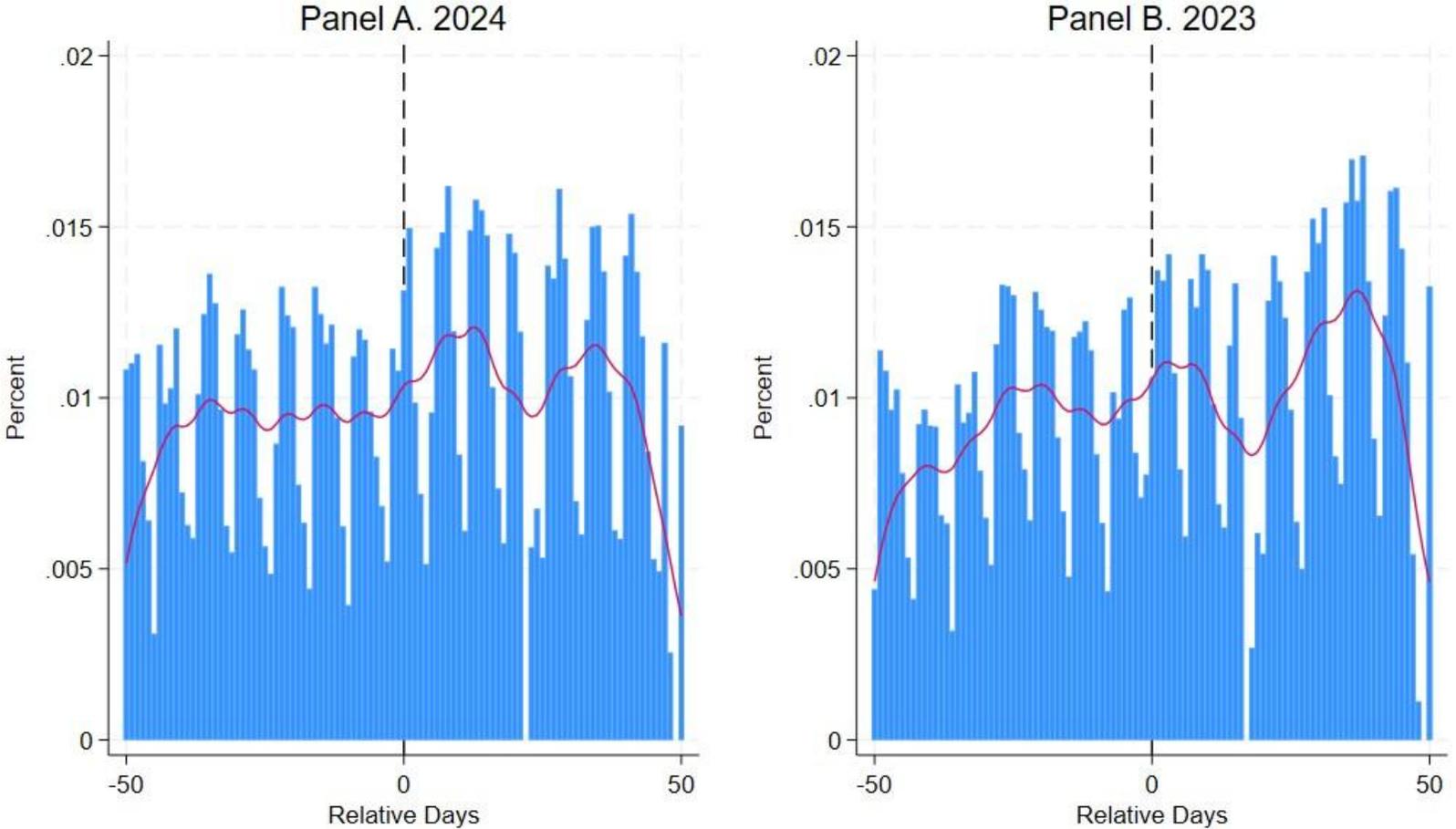

Figure A.1: Sample Distributions in 2024 and 2023

*Note*: The left figure displays the distribution of the number of observations of the analysis sample. The right figure displays the distribution of the number of observations of the placebo test sample. The 2023 and 2024 waves of the Behavioral Risk Factor Surveillance System are used for the figure.

Table A.1: Analysis Sample vs Entire Sample

| | (1) | (2) | (3) | (4) |
|---|---|---|---|---|
| | Analysis Sample | | Entire Sample | |
| | Mean | Std Dev | Mean | Std Dev |
| *Panel A. Outcome Variable* | | | | |
| Bad Mental Health | 0.402 | 0.490 | 0.395 | 0.489 |
| Very Bad Mental Health | 0.134 | 0.341 | 0.132 | 0.338 |
| Satisfied with Life | 0.944 | 0.230 | 0.945 | 0.229 |
| Very Satisfied with Life | 0.450 | 0.497 | 0.452 | 0.498 |
| *Panel B. Covariates* | | | | |
| Male | 0.476 | 0.499 | 0.474 | 0.499 |
| Age 18 to 24 | 0.056 | 0.230 | 0.058 | 0.233 |
| Age 25 to 34 | 0.099 | 0.298 | 0.099 | 0.299 |
| Age 35 to 44 | 0.125 | 0.331 | 0.128 | 0.334 |
| Age 45 to 54 | 0.137 | 0.344 | 0.136 | 0.342 |
| Age 55 to 64 | 0.173 | 0.379 | 0.176 | 0.381 |
| Age 65 and over | 0.410 | 0.492 | 0.403 | 0.491 |
| Married | 0.524 | 0.499 | 0.524 | 0.499 |
| High School Diploma | 0.949 | 0.219 | 0.948 | 0.221 |
| College Diploma | 0.434 | 0.496 | 0.436 | 0.496 |
| Employed | 0.498 | 0.500 | 0.499 | 0.500 |
| White | 0.765 | 0.424 | 0.755 | 0.430 |
| Urban | 0.861 | 0.346 | 0.868 | 0.339 |
| Sample Size | 51,740 | | 190,488 | |

Note: This table presents the descriptive statistics of the analysis sample and the full 2024 wave of the Behavioral Risk Factor Surveillance System (BRFSS). Columns (1) and (2) show the means and standard deviations for the analysis sample, while columns (3) and (4) report the means and standard deviations for the full 2024 wave of the BRFSS.

Table A.2: Heterogeneity by Age, Marital Status, Education Level, and Employment Status

|  | (1) | (2) | (3) | (4) | (5) | (6) | (7) | (8) |
|---|---|---|---|---|---|---|---|---|
|  | Age < 55 | Age ≥ 55 | Married | Not Married | No College | College Diploma | Employed | Not Employed |
| *Panel A. Bad Mental Health* | | | | | | | | |
| Election | 0.048*** | 0.037** | 0.051*** | 0.033* | 0.024 | 0.062*** | 0.054*** | 0.030* |
|  | (0.017) | (0.019) | (0.016) | (0.019) | (0.017) | (0.018) | (0.016) | (0.017) |
| *Panel B. Very Bad Mental Health* | | | | | | | | |
| Election | 0.027** | 0.007 | 0.006 | 0.025** | 0.018 | 0.007 | 0.018* | 0.015 |
|  | (0.011) | (0.009) | (0.010) | (0.011) | (0.011) | (0.012) | (0.010) | (0.011) |
| *Panel C. Satisfied with Life* | | | | | | | | |
| Election | -0.024** | -0.003 | -0.010* | -0.011 | -0.013 | -0.008 | -0.013* | -0.010 |
|  | (0.009) | (0.007) | (0.005) | (0.010) | (0.009) | (0.007) | (0.007) | (0.008) |
| *Panel D. Very Satisfied with Life* | | | | | | | | |
| Election | 0.009 | -0.051*** | -0.037** | -0.017 | -0.045*** | -0.007 | -0.020 | -0.036*** |
|  | (0.021) | (0.017) | (0.018) | (0.015) | (0.017) | (0.021) | (0.020) | (0.014) |
| Sample Size | 21,578 | 30,162 | 27,094 | 24,646 | 29,265 | 22,475 | 25,758 | 25,982 |

Note: This table displays subgroup analyses by each group with the full set of covariates which include age, gender, and race, marital status, education, employment status, and the state fixed effects. Standard errors in parentheses are clustered at the state level. The 2024 wave of the Behavioral Risk Factor Surveillance System is used for the figure. *, **, and *** denote statistical significance at the 90 percent, 95 percent, and 99 percent levels, respectively.